\documentclass[aps,prl,twocolumn,superscriptaddress,longbibliography,10pt]{revtex4-2}

\usepackage{amsmath,amssymb,amsbsy,amstext,amsfonts,mathrsfs,latexsym,bm,physics,graphicx,xcolor}
\usepackage[colorlinks=true,urlcolor=blue,anchorcolor=blue,citecolor=blue,linkcolor=blue,linktocpage=true,pdfa=true]{hyperref}
\usepackage{orcidlink}

\def\sqrtb{\mathpalette\DHLhksqrt}
\def\DHLhksqrt#1#2{%
\setbox0=\hbox{$#1\sqrt{#2\,}$}\dimen0=\ht0
\advance\dimen0-0.2\ht0
\setbox2=\hbox{\vrule height\ht0 depth -\dimen0}%
{\box0\lower0.4pt\box2}}

\newcommand{\mpl}{M_{\scriptscriptstyle\mathrm{Pl}}}
\newcommand{\lth}{\lambda_{\scriptscriptstyle\mathrm{tH}}}

\begin{document}

\title{Ultraviolet Completion of the Big Bang in Quadratic Gravity}

\author{Ruolin Liu\,\orcidlink{0009-0003-7385-3071}}
\affiliation{Department of Physics and Astronomy, University of Waterloo, Waterloo, ON N2L 3G1, Canada}
\affiliation{Perimeter Institute for Theoretical Physics, Waterloo, ON N2L 2Y5, Canada}
\affiliation{Waterloo Centre for Astrophysics, University of Waterloo, Waterloo, ON N2L 3G1, Canada}
\author{Jerome Quintin\,\orcidlink{0000-0003-4532-7026}}
\affiliation{École de technologie supérieure, Université du Québec, Montr\'eal, QC H3C 1K3, Canada}
\affiliation{Department of Physics, McGill University, Montr\'eal, QC H3A 2T8, Canada}
\affiliation{Department of Applied Mathematics, University of Waterloo, Waterloo, ON N2L 3G1, Canada}
\affiliation{Perimeter Institute for Theoretical Physics, Waterloo, ON N2L 2Y5, Canada}
\affiliation{Waterloo Centre for Astrophysics, University of Waterloo, Waterloo, ON N2L 3G1, Canada}
\author{Niayesh Afshordi\,\orcidlink{0000-0002-9940-7040}}
\affiliation{Department of Physics and Astronomy, University of Waterloo, Waterloo, ON N2L 3G1, Canada}
\affiliation{Perimeter Institute for Theoretical Physics, Waterloo, ON N2L 2Y5, Canada}
\affiliation{Waterloo Centre for Astrophysics, University of Waterloo, Waterloo, ON N2L 3G1, Canada}

\begin{abstract}
We present a quantum quadratic gravity inflationary scenario that can accommodate the new cosmological constraints, which have disfavored Starobinsky inflation. The theory is asymptotically free in the ultraviolet, but 1-loop running is found to dynamically lead to slow-roll inflation toward the infrared. When a large number of matter fields contribute to the beta functions, the spectral index and the tensor-to-scalar ratio can be phenomenologically viable. We find that as inflation ends, the theory approaches its strong coupling regime and general relativity must emerge, as an effective field theory, as the universe must reheat and enter its standard radiation era. In order to avoid strong coupling, a minimum tensor-to-scalar ratio of 0.01 is predicted for this theory. Our framework offers a laboratory for connecting a concrete ultraviolet completion (quantum quadratic gravity) with inflationary dynamics, reheating, and precise cosmological observations.
\end{abstract}

\maketitle


General Relativity (GR) is tremendously successful when treated as an effective field theory (EFT) \cite{Burgess:2003jk}. This means, though, that there is a cutoff scale beyond which it cannot be trusted. Beyond this scale, one has to deal with GR's nonrenormalizability and with singularities that develop in solutions to the classical Einstein equations. When starting with the linear-in-curvature Einstein-Hilbert (EH) action \footnote{We use the mostly positive metric signature. We contract tensor indices with the metric $g_{\mu\nu}$, whose determinant is $g$ and Ricci scalar $R$. Finally, we use natural units, and $\mpl$ denotes the reduced Planck mass.}, $S_{\rm EH}=\frac{1}{2}\int\dd^4x\sqrt{-g}\mpl^2R$, one has to include an infinite hierarchy of higher curvature counterterms to cancel divergences in large-momentum loops \cite{tHooft:1974toh,Christensen:1979iy,Goroff:1985th}, but this is not the case if the bare action is augmented only by terms that are quadratic in curvature \cite{Stelle:1976gc}. As such, including quadratic gravity,
\begin{equation}
    S_{\rm quadratic~gravity} = -\int\dd^4x\,\sqrt{-g}\left(\frac{R^2}{\xi}+\frac{C^2}{2\lambda}\right)\,,
\end{equation}
allows for a possible ultraviolet (UV) completion of GR. It is thus compelling to explore UV regimes where the effects of quadratic gravity could manifest themselves and be tested, such as in black holes \cite{Holdom:2016nek,Liu:2024zti} and in the early universe (e.g., \cite{Salvio:2018crh}). This \textit{letter} explores the latter.

When adding the square of the Ricci scalar ($R^2$) to the EH action, one already has the starting point of Starobinsky inflation \cite{Starobinsky:1980te}. However, in general, quadratic gravity also contains the contraction of the Weyl tensor with itself ($C^2$) \footnote{The action of quadratic gravity can generically be written as such up to boundary and topological terms (see, e.g., \cite{Salvio:2018crh}).}. Hence, as a higher-derivative theory, the theory comes with two new fields \cite{Hinterbichler:2015soa,Salvio:2018crh} in addition to the massless graviton in GR: a massive spin-0 field coming from $R^2$ (which can act as the inflaton) and a massive spin-2 field coming from $C^2$. The latter field is a ghost; i.e., it carries negative kinetic energy, leading to an unbounded Hamiltonian (if we ignore interactions). There is a lot of literature attempting to make sense of ghosts in various settings \cite{Hawking:2001yt,Mannheim:2004qz,Mannheim:2006rd,Bender:2007wu,Donoghue:2017fvm,Donoghue:2018izj,Donoghue:2019fcb,Donoghue:2019ecz,Donoghue:2020mdd,Donoghue:2021eto,Donoghue:2021meq,Donoghue:2021cza,Salvio:2018crh,Edelstein:2021jyu,Hell:2023rbf,Holdom:2023usn,Holdom:2024cfq,Holdom:2024onr,Edelstein:2024jzu,Salvio:2024joi,Lambiase:2025qyl}, which we do not review here, but in the context of inflationary cosmology different techniques allow one to handle the ghost, its quantization, and the corresponding analysis of Starobinsky-like inflation, where the presence of $C^2$ only leads to (small) corrections to the tensor perturbations \cite{Clunan:2009er,Deruelle:2010kf,Deruelle:2012xv,Anselmi:2020lpp,DeFelice:2023psw,Kubo:2025jla,Bianchi:2025tyl,Buoninfante:2025dgy}.

Instead, here we wish to explore another possibility --- arguably more radical --- distinct from Starobinsky's $R+R^2$ inflation. As Quantum Quadratic Gravity (QQG) is UV complete, one may start at the `Big Bang' singularity at infinite curvature. There, QQG is `pure' --- there is no GR. In this regime, the theory is amenable to a perturbative treatment, and accordingly, beta functions at the 1-loop order reveal asymptotic freedom in the UV \cite{Fradkin:1981iu,Avramidi:1985ki,Codello:2006in,Niedermaier:2009zz,Niedermaier:2010zz,Ohta:2013uca}. QQG has many similarities to quantum chromodynamics (QCD) and suggests that GR may emerge as QQG becomes strongly coupled in the infrared (IR) \cite{Holdom:2015kbf,Holdom:2016xfn} (see also the recent proposal \cite{deBoer:2025vpx}); from the low energy perspective, QQG would be confined. The question thus becomes: can one find a successful early universe scenario in the deep UV, solely from QQG, i.e., quadratic gravity with running taken into account?

We should highlight that what would make this scenario unique is that gravity does become strongly coupled at some scale, which is also used to confine ghost degrees of freedom \cite{Holdom:2015kbf,Holdom:2016xfn}. In contrast, in Starobinsky inflation, gravity remains weakly coupled, and thus the containment of ghosts and Ostrogradsky instabilities would require a more creative solution.

The renormalization group (RG) flow of QQG and other higher-derivative theories has been the subject of renewed interest in recent years \cite{Buccio:2024hys,Kawai:2024aim,Buccio:2024omv,Kawai:2025wkp,Salvio:2025cmi}. In particular, \cite{Buccio:2024hys} argued that the previously derived 1-loop beta functions of QQG did not capture the physical scale at which interactions take place. New beta functions have been computed,
\begin{align}
    \beta_\xi&=\frac{\dd\xi}{\dd\ln\mu}=-\frac{1}{(4\pi)^2}\frac{\xi^2-36\lambda\xi-2520\lambda^2}{36}\,,\nonumber\\
    \beta_\lambda&=\frac{\dd\lambda}{\dd\ln\mu}=-\frac{1}{(4\pi)^2}\frac{\big[\big(1617+90\mathcal{N}\big)\lambda-20\xi\big]\lambda}{90}\,,\label{eq:betaFuncFullWithMatter}
\end{align}
where $\mu$ represents here the \emph{physical} running scale of the QQG couplings $\xi$ and $\lambda$. Although the validity of these `physical' beta functions is still a matter of active debate \cite{Kawai:2024aim,Buccio:2024omv,Kawai:2025wkp,Salvio:2025cmi}, in this {\it letter} we shall argue that this formulation [in particular, the emergence of a maximum for $\xi(\mu)$] would allow for a successful inflationary scenario. Thus, we use this observation as empirical guidance that may inform the ongoing theoretical debate.   

\begin{figure*}[ht]
    \centering
    \includegraphics[width=\textwidth]{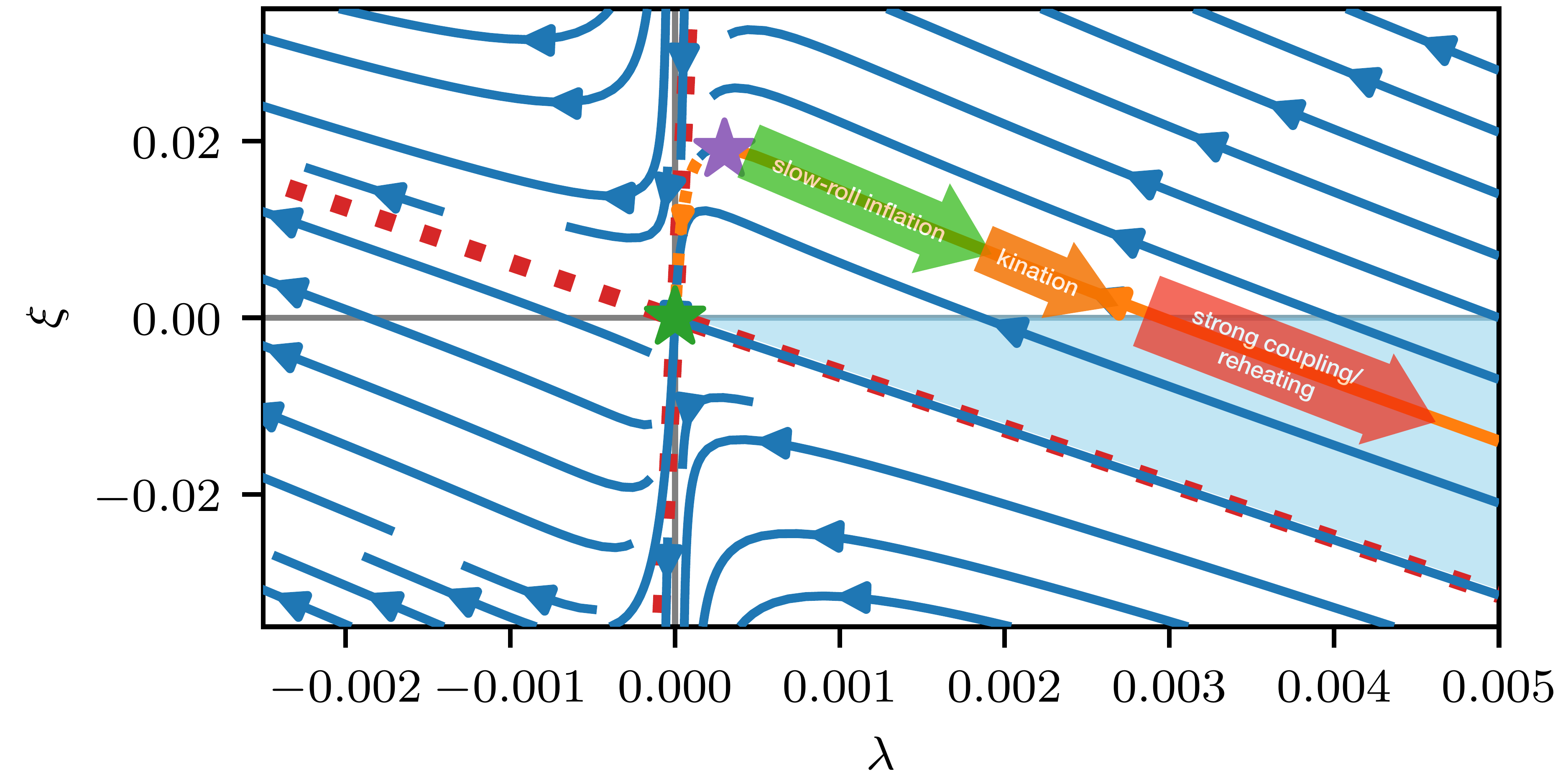}
    \caption{The renormalization group (RG) flow of quadratic gravity with $\mathcal{N}=10$ matter-field content. The two red dashed lines show the separatrices of the RG flow, while the light blue shaded region is tachyon-free \cite{Buccio:2024hys}. The orange trajectory shows a candidate cosmological evolution: starting from a potential tunneling from a (no-boundary) Euclidean geometry at the top of the trajectory (purple star) $\rightarrow$ slow-roll inflation (potential-dominated) $\rightarrow$ kination (kinetic-dominated) $\rightarrow$ strong-coupling/reheating, where an EH term with effective $\mpl$ (i.e., GR) emerges eventually.}
    \label{fig:streamplot}
\end{figure*}

While \cite{Buccio:2024hys} derived these equations in vacuum, we added the possible contribution from matter fields in the loops; the number of such fields,
\begin{equation}
    \mathcal{N}=\frac{1}{60}\mathcal{N}_\mathrm{scalar}+\frac{1}{5}\mathcal{N}_\mathrm{vector}+\frac{1}{20}\mathcal{N}_\mathrm{fermion}\,,
\end{equation}
correspondingly enhances $\beta_\lambda$ \cite{Salvio:2018crh}. Exact analytic (though implicit) solutions can be derived in the vacuum ($\mathcal{N}=0$) case \setcounter{footnote}{98}\footnote{See the Supplemental Material.}. The general features remain the same even as $\mathcal{N}$ is increased; we show the case of $\mathcal{N}=10$ in Fig.~\ref{fig:streamplot} for illustrative purposes. The key aspect of the results of \cite{Buccio:2024hys} is that the beta functions now admit solutions that are both asymptotically free in the UV (reaching the green star at the origin in Fig.~\ref{fig:streamplot}) and tachyon free in the IR, which is to say that low-$\mu$ basin of attraction is where $\lambda>0$ and $\xi<0$ (the pale blue region in Fig.~\ref{fig:streamplot}); the orange curve is an example of such a trajectory. Starobinsky inflation could happen precisely in the tachyon-free region, should one include an EH term in the action (see \cite{Percacci:2025ehx}).

The scenario we wish to put forward is different, as we wish to start with pure QQG in the UV. Looking at trajectories that would eventually reach the tachyon-free region (where we would expect GR to emerge), we see that all of them (except the one trajectory that follows the red dotted separatrix) must first go in the region where both $\lambda$ and $\xi$ are positive: $\lambda$ always monotonically grows, while $\xi$ first grows before decreasing and eventually crossing the tachyon divide at a scale that we call $\mu_0$. What we find is that a successful inflationary phase can be realized in the regime where $\xi$ is decreasing, though before it crosses the tachyon divide. We find that as inflation ends, the universe enters a kinetic-dominated phase (kination) and QQG approaches its strong coupling scale. Eventually, GR must emerge, and the universe must enter its radiation era. One can thus think of the reheating surface at the end of kination following inflation as a UV/IR matching surface for QQG/GR.

Here is how inflation arises in pure QQG: Considering a homogeneous and isotropic background, there is no contribution from the Weyl tensor, so the action is solely that of $R^2$ gravity. Without running, the theory is scale invariant \cite{Hell:2023mph,Hell:2025wha}, and any constant-$R$ spacetime extremizes the action, such as Minkowski and (anti-)de Sitter. The exact scale invariance is broken once a physical scale appears in the theory, such as $\mpl$ if one includes GR; that is how one goes from de Sitter to quasi-de Sitter in Starobinsky inflation. However, even without such explicit scale, something similar occurs if we consider the running due to quantum effects: the theory is not exactly pure $R^2$ gravity anymore, and scale invariance is broken, akin to the so-called quantum {\it conformal anomaly} \cite{Capper:1974ic,Duff:1993wm}.

While the onset of inflation in this scenario (like all the alternatives) remains speculative, a natural possibility is a no-boundary Euclidean manifold \cite{Hartle:1983ai,Vilenkin:1982de}. Such a manifold would be an exact solution to the Euclidean QQG at the maximum of $\xi$ (where its derivative vanishes, depicted by the purple star in Fig.~\ref{fig:streamplot}), as the Euclidean action for a 4-sphere is $S_\mathrm{E} \propto \xi^{-1}$ in this theory. Across its equator, the Euclidean half-sphere can then be matched to the waist of a closed (Lorentzian) de Sitter spacetime, which would describe cosmology when $t \to -\infty$. However, this de Sitter phase is unstable due to RG running of $\xi$, which is expected to initiate slow-roll inflation \footnote{The natural possibility of a `creation out of nothing' is a compelling one here, but it certainly deserves further investigation. In particular, finding proper instanton solutions and analyzing the path integral of QQG would constitute interesting follow-up work.}.

To explore the effects of RG flow, we start with one important assumption, which is that we are not in vacuum, but that instead a large number of matter fields are present. This is usually a fair assumption in most models that go beyond the standard model of particle physics \footnote{One may question how large $\mathcal{N}$ can reasonably be. In theoretical settings, such as holography, $\mathcal{N}$ is often taken to be arbitrarily large, to ensure perturbative gravity in the bulk. Here, we remain agnostic about whether or not a very large $\mathcal{N}$ represents a fine-tuning problem. We should further note that these matter fields would probably end up being confined to the UV, together with all the degrees of freedom of QQG, in a similar fashion to quarks and gluons in QCD.}. This is not to say that such matter fields are necessarily excited at the Big Bang; in fact, we assume that they sit in their vacua. Still, vacuum fluctuations of $\mathcal{N}\gg 1$ fields have important implications in the running of QQG, and thus indirectly, these matter fields contribute to modifying the cosmological background and perturbations. The solution to the beta functions \eqref{eq:betaFuncFullWithMatter} in the large-$\mathcal{N}$ limit can be expressed as \cite{Note99}
\begin{equation}
    \xi(\mu)\simeq\frac{35\lambda_0^2\ln(\mu/\mu_0)}{8\pi^2[1+\lth \ln(\mu/\mu_0)]}\,,\label{eq:xiofmulargeN}
\end{equation}
where $\mu_0$ is the scale where $\xi=0$ and $\lambda=\lambda_0$ \footnote{Here, we ignore the impact of the running of the coupling to Gauss-Bonnet's invariant $\mathcal{G}$, which is not widely studied due to its topological nature. However, this may affect cosmological evolution --- as in $f(R,\mathcal{G})$ theories, which are widely studied in the literature, e.g., \cite{Cognola:2006eg} --- but we defer that consideration to future work.}. Here, we define a 't~Hooft-like coupling constant, $\lth \equiv\lambda_0\mathcal{N}/(4\pi)^2$, which is the key quantity in assessing the strength of the loop corrections when $\mathcal{N}\gg 1$; the 1-loop approximation is under control so long as $\lambda\lesssim\lambda_0$ and $\lth\lesssim 1$, so $\lambda_0$ must be sufficiently small at large $\mathcal{N}$. We shall see that as $\lambda$ flows beyond $\lambda_0$ (as we cross the tachyon divide), strong coupling must be reached and GR must emerge as $\lambda\mathcal{N}/(4\pi)^2$ surpasses unity.

At this point, different techniques can be employed to compute the quantum effective action \cite{Buchbinder:1992rb} that results from the running of the couplings $\xi$ and $\lambda$. We take the simple approach of promoting $\mu$ to a covariant definition of energy scale and substituting the resulting coupling into the action. This approach is ambiguous, though, to the extent that it depends on the choice of energy scale for $\mu$, but there are many examples (including in inflation; e.g., \cite{Elizalde:1993ee,Elizalde:1993qh,Myrzakulov:2014hca,Glavan:2023lvw}) where it is reasonable to expect the RG scale to correspond to the background cosmological/curvature scale. As such, many curvature invariants could be chosen, e.g., the Kretschmann scalar, but we take what is possibly the simplest: the Ricci scalar \footnote{Note that during the slow-roll phase of inflation, due to the near de-Sitter nature of spacetime, different choices of curvature invariants for the RG scale are generically equivalent to the Ricci scalar. However, they will start to differ as we approach the end of inflation, and thus, they could lead to different reheating mechanisms in the strongly coupled regime.}. Considering the action of QQG without the Weyl tensor first \footnote{The contribution from the Weyl tensor vanishes on a homogeneous and isotropic background. This is not true once cosmological perturbations are introduced. Still, contributions from the Weyl tensor to, e.g., tensor perturbations are usually highly suppressed \cite{Deruelle:2010kf}, and we do not expect this result to be affected by small logarithmic running.}, this means that we can write the action as an $f(R)$ theory of gravity; taking the solution \eqref{eq:xiofmulargeN}, it is $f(R)=R^2/\xi(R)=8\pi^2(\lth +4/\ln[R^2/\mu_0^4])R^2/(35\lambda_0^2)$. The RG flow thus yields a small (logarithmic) correction to the pure $R^2$ action, which leads to quasi-de Sitter attractor solutions \cite{Note99}.

As observables at this level should not depend on the choice of frame (e.g., \cite{Anselmi:2020lpp}), let us transform to the Einstein frame, where more intuition can be gained since the $f(R)$ theory becomes equivalent to GR with a scalar field. Sufficiently far from the end of inflation, the scalar field's potential can be approximated as \cite{Note99}
\begin{equation}
    V(\varphi)\simeq\frac{35\lambda_0^2\mu_0^4}{128\pi^2\lth }\left(1-\frac{\sqrtb{6}\mu_0}{\lth \varphi}\right)\,.\label{eq:Vf}
\end{equation}
This falls in the brane inflation generalized \emph{phenomenological} classification of \cite{Martin:2013tda}, but as far as we are aware, it is the first time that this specific form of the potential is derived from a UV-complete theory. The potential approximated as \eqref{eq:Vf} only holds at early times, but it nevertheless still captures an important finding about late times \cite{Note99}, which is that inflation has a graceful exit mechanism as the potential steepens, but also that it does not reach a minimum where the inflaton could decay. Instead, the field rolls ever faster into a kinetic-dominated phase (kination). Eventually, though, nonperturbative effects would alter the dynamics as the theory enters its strong coupling regime.

Assuming that the number of $e$-folds $N$ in between the end of inflation and the time of horizon exit for modes of cosmological interest is sufficiently large (50 to 60, similar to other inflationary models \cite{Planck:2013jfk}; see \cite{Note99} for an assessment of this assumption), we find the amplitude of the scalar perturbations to be given by $A_\mathrm{s}\sim 35\lambda_0^2N^{4/3}/[512\pi^4(2\lth )^{1/3}]$, while the scalar spectral index and the tensor-to-scalar ratio are, respectively, \cite{Note99}
\begin{equation}
    n_\mathrm{s}\sim 1-\frac{4}{3N}\,,\qquad r\sim\frac{8}{3}\left(\frac{2}{\lth^2N^4}\right)^{1/3}\,.
\end{equation}
This somewhat resembles the predictions of Starobinsky inflation, though both are larger than Starobinsky's $n_\mathrm{s}\approx 1-2/N$ and $r\approx 12/N^2$.

\begin{figure}[th]
    \centering
    \includegraphics[width=0.98\linewidth]{nsrplot.png}
    \caption{Plot of the scalar spectral index $n_\mathrm{s}$ vs the tensor-to-scalar ratio $r$ for Starobinsky inflation (in black), contrasted with our QQG model in color for different values of the coupling $\lth$. The purple and dashed contours correspond to a combination of CMB (\textit{Planck}18+ACT+SPT+Lensing+BK) and BAO (DESI) observational constraints for $\Lambda$CDM and $w_0w_a$CDM cosmologies, respectively \cite{SPT-3G:2025bzu}.}
    \label{fig:nsr}
\end{figure}

Using the exact expressions derived in \cite{Note99}, we can plot the predictions in the $n_\mathrm{s}$-$r$ plane (see Fig.~\ref{fig:nsr}), where this is also contrasted with Starobinsky inflation. Dots and lines of different color correspond to different values of the 't~Hooft-like coupling $\lth$ in QQG. We can see that $\lth$ has little effect on $n_\mathrm{s}$, but the more weakly coupled the theory is (so the smaller $\lth$ is), the larger $r$. In comparison, we show a combination of constraints from cosmic microwave background (CMB) data (from \textit{Planck} \cite{Planck:2018jri}, ACT \cite{AtacamaCosmologyTelescope:2025blo}, SPT \cite{SPT-3G:2025bzu}, and BICEP/\textit{Keck} \cite[BK]{BICEP:2021xfz}) and baryon acoustic oscillation (BAO) data (from DESI \cite{DESI:2024mwx}). Such data combination (especially ACT with DESI within $\Lambda$CDM; see \cite{Ferreira:2025lrd} for a discussion of the tension) prefers larger $n_\mathrm{s}$ values, putting Starobinsky inflation in slight tension with observations but placing our QQG model in a favorable position. Nonetheless, allowing for a dynamical dark energy \footnote{Cold dark matter with a cosmological constant ($\Lambda$CDM) refers to the standard model of cosmology, while the Chevallier-Polarski-Linder \cite{Chevallier:2000qy,Linder:2002et} parametrization for dynamical dark energy (often called $w_0w_a$CDM) stands as an alternative where the dark-energy equation of state depends linearly on the cosmological scale factor. The recent BAO data of DESI point toward a slight preference for the latter \cite{DESI:2024mwx}.} would put both Starobinsky inflation and QQG within 1$\sigma$ of observations \cite{SPT-3G:2025bzu}. 

\begin{figure}[th]
    \centering
    \includegraphics[width=0.9\linewidth]{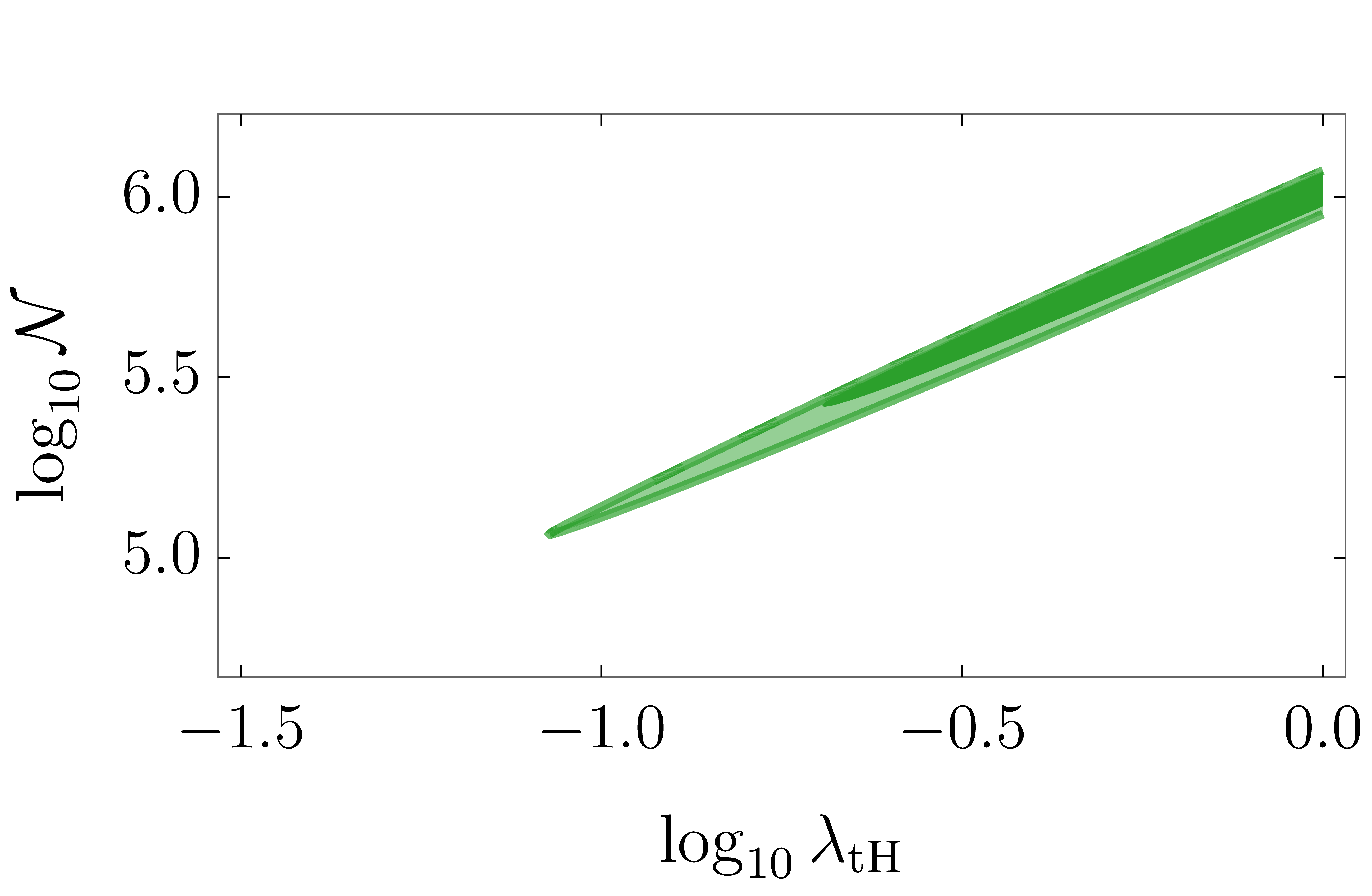}
    \caption{Constraints on the QQG parameters $\lth$ and $\mathcal{N}$ from the observationals bounds on $n_\mathrm{s}$, $r$, and $A_\mathrm{s}$ combined. The dark and pale green regions amount to $1\sigma$ and $2\sigma$ confidence intervals, respectively.}
    \label{fig:tHvcalNconstr}
\end{figure}

Given a fixed number of inflationary $e$-folds, the observables $n_\mathrm{s}$, $r$, and $A_\mathrm{s}$ solely depend on $N$, $\lambda_0$, and $\lth\equiv\lambda_0\mathcal{N}/(4\pi)^2$. Thus, observational constraints on those quantities can be translated into bounds on $\lth$ and $\mathcal{N}$. We present such constraints in Fig.~\ref{fig:tHvcalNconstr}, where we set the scalar amplitude to $A_\mathrm{s}=e^{3.06}\times 10^{-10}$ \cite{Planck:2018jri}, and $1\sigma$ and $2\sigma$ confidence intervals on $n_\mathrm{s}$ and $r$ are taken as in Fig.~\ref{fig:nsr}, while for the $e$-folding number we take $50\pm 10$. Since $r\sim \lth^{-2/3}$, we can see in Fig.~\ref{fig:tHvcalNconstr} that there is no upper bound on $\lth$ because there is no observational lower bound on $r$. However, we do not trust the results beyond $\lth\gtrsim 1$ (in the strong coupling regime), hence the cutoff at $\lth=1$. As such, the viable parameter space is spanned by $0.1\lesssim\lth\lesssim 1$ and $10^5\lesssim\mathcal{N}\lesssim 10^6$. It thus appears that two things are phenomenologically preferred: that there must be a (very) large number of matter fields present in the theory; and that the theory must be very close to strong coupling as $\lambda$ approaches the tachyon divide at $\lambda_0$ since $\lth\equiv\lambda_0\mathcal{N}/(4\pi)^2$ is likely very close to unity. The latter brings us back full circle to our earlier claim: current observations suggest that crossing the tachyon divide appears to be nearly coincident with entering the strong-coupling regime, $\lth \gtrsim 1$, as well as with the onset of reheating (red arrow in Fig.~\ref{fig:streamplot}).


Let us summarize our findings. We presented a UV-complete `quantum quadratic gravity' inflationary scenario that can be compatible with the recent CMB constraints, which may otherwise disfavor the standard Starobinsky $R+R^2$ inflation. By identifying the physical renormalization scale with curvature, $\mu=|R|^{1/2}$, pure $R^2$ gravity acquires controlled logarithmic corrections and yields an almost-plateau potential in the Einstein frame. This quantum quadratic gravity preserves the slow-roll virtues of plateau models while shifting predictions toward the parameter space preferred by recent CMB observations. In particular, for a large number of matter fields, $\mathcal{N}\sim\mathcal{O}(10^{5}-10^{6})$, the 't~Hooft-like coupling $\lambda_{\rm tH}$ can stay under perturbative control, and the predicted $\{n_\mathrm{s},r,A_\mathrm{s}\}$ are consistent with the data. In order to avoid strong coupling, the tensor-to-scalar ratio, $r$, is predicted to be $\gtrsim 0.01$.

Many things remain to be explored. First, incorporating two-loop RG equations for the quadratic couplings would test the robustness of the quantum inflation potential (especially as we approach $\lth \sim 1$), refine the viable parameter space, and quantify scheme/threshold uncertainties as GR effectively emerges. The RG flow of other couplings, such as Gauss-Bonnet, Yang-Mills, and Yukawa, may lead to richer phenomenology as observations improve (e.g., \cite{Afshordi:2016dvb}). Second, while the Weyl term $C^2$ vanishes on homogeneous and isotropic backgrounds, it affects perturbations; assessing its impact --- together with possible ghost mechanisms and the ensuing stability/observables --- is an important next step. One can also think of comparing this RG-improved setup with holographic cosmology (which also requires loop contributions with a large value of $\mathcal{N}$ \cite{McFadden:2009fg,Afshordi:2016dvb}) and the Hartle-Hawking no-boundary proposal, clarifying the role of initial state in our QQG inflation. Moreover, a kinetic-dominated (kination) phase naturally appears at the end of inflation; analyzing the onset of reheating, the strong-coupling window, and the EFT handover to GR could sharpen the link to data. Finally, beyond ACT, a combined analysis with \textit{Planck}/BICEP/SPT/Simons Observatory could probe the reported ACT tension~\cite{Ferreira:2025lrd} and forecast $r$ within next-generation sensitivities. Overall, this framework offers a concrete laboratory connecting RG running in renormalizable quadratic gravity to inflationary dynamics, reheating, and precision CMB data, while inviting broader comparisons with UV-complete cosmological scenarios.


\paragraph{Acknowledgments---}We would like to thank Luca Buoninfante, Dra\v{z}en Glavan, Anamaria Hell, Roberto Percacci, and Kostas Skenderis for useful discussions.
This research was supported in part by the Perimeter Institute for Theoretical Physics and the National Sciences and Engineering Research Council of Canada (NSERC).
Research at the Perimeter Institute is supported by the Government of Canada through the Department of Innovation, Science and Economic Development and by the Province of Ontario through the Ministry of Colleges and Universities.
J.Q.~further acknowledges financial support from the University of Waterloo's Faculty of Mathematics William T.~Tutte Postdoctoral Fellowship during his time at UWaterloo.


\bibliographystyle{apsrev4-2}
\bibliography{refs}


\appendix
\onecolumngrid
\newpage

\section{Supplemental Material}
\setcounter{page}{1}
\setcounter{equation}{0}

\subsection{Exact and approximate solutions of the renormalization group equations}

\subsubsection{In vacuum}

The beta functions of \cite{Buccio:2024hys} in vacuum,
\begin{align}
        \beta_\xi&=\frac{\dd\xi}{\dd\ln\mu}=-\frac{1}{(4\pi)^2}\frac{\xi^2-36\lambda\xi-2520\lambda^2}{36}\,,\nonumber\\
        \beta_\lambda&=\frac{\dd\lambda}{\dd\ln\mu}=-\frac{1}{(4\pi)^2}\frac{\left(1617\lambda-20\xi\right)\lambda}{90}\,,\label{eq:Buccio}
\end{align}
admit exact (implicit) solutions. It is easier to describe those solutions by defining the ratio $\tilde\omega\equiv -15\xi/\lambda$. Setting the constants $\mu_0$ and $\lambda_0$ such that $\lambda(\mu_0)=\lambda_0$ and $\xi(\mu_0)=0$, one finds
\begin{align}
    &\lambda(\mu)=\lambda_0\left(1-\frac{\tilde\omega(\mu)}{\Delta_-}\right)^{-(4+p)/9}\left(1+\frac{\tilde\omega(\mu)}{\Delta_+}\right)^{-(4-p)/9}\,,\nonumber\\
    &A+\frac{\lambda_0}{B}\ln\left(\frac{\mu}{\mu_0}\right)=\left[\tilde\omega(\mu)+\Delta_+\right]^{(4-p)/9}{}_2F_1\left(\frac{4-p}{9},\frac{5-p}{9};\frac{13-p}{9};\frac{\tilde\omega(\mu)+\Delta_+}{2\theta}\right)\,,
\end{align}
where ${}_2F_1$ denotes the hypergeometric function.
The numbers defined in the above are $\theta\equiv\sqrt{386\,761}$, $\Delta_\pm\equiv\theta\pm 569$, $p\equiv 2575/\theta$, and
\begin{align}
    &B\equiv\frac{9\pi^2\cdot 3^{1/9}\cdot 2^{(37+p)/9}\theta^{-(5-p)/9}}{5^{1/3}\cdot 7^{4/9}(p-4)}\left(\frac{\Delta_+}{\Delta_-}\right)^{p/9}\,,\nonumber\\
    &A\equiv\Delta_+^{(4-p)/9}{}_2F_1\left(\frac{4-p}{9},\frac{5-p}{9};\frac{13-p}{9};\frac{\Delta_+}{2\theta}\right)\,.
\end{align}
If one expands the above about $\tilde\omega\approx 0$ (so $|\xi|\ll|\lambda|$) and further assumes $|\lambda_0|\ll 1$, one can asymptotically solve the implicit solutions as
\begin{align}
    \lambda(\mu)&\simeq\lambda_0-\frac{539}{480}\frac{\lambda_0^2}{\pi^2}\ln\left(\frac{\mu}{\mu_0}\right)+\frac{297\,521}{230\,400}\frac{\lambda_0^3}{\pi^4}\ln^2\left(\frac{\mu}{\mu_0}\right)\,,\nonumber\\
    \xi(\mu)&\simeq\frac{35}{8}\frac{\lambda_0^2}{\pi^2}\ln\left(\frac{\mu}{\mu_0}\right)-\frac{917}{192}\frac{\lambda_0^3}{\pi^4}\ln^2\left(\frac{\mu}{\mu_0}\right)\,.\label{eq:approxSolBetaVac}
\end{align}
One finds the same thing by solving the system \eqref{eq:Buccio} of ordinary differential equations order by order about $\lambda\approx\lambda_0$ and $\xi\approx 0$.

\subsubsection{With matter fields}

Upon considering matter fields and their contributions in loops, one expects $\beta_\lambda$ to receive a correction term as given in \eqref{eq:betaFuncFullWithMatter}. In the limit where $\mathcal{N}\gg 539/30$ and $\lambda\mathcal{N}\gg 2\xi/9$, the beta functions reduce to
\begin{align}
    \beta_\xi&=\frac{\dd\xi}{\dd\ln\mu}=-\frac{1}{(4\pi)^2}\frac{\xi^2-36\lambda\xi-2520\lambda^2}{36}\,,\nonumber\\
    \beta_\lambda&=\frac{\dd\lambda}{\dd\ln\mu}\simeq-\frac{1}{(4\pi)^2}\mathcal{N}\lambda^2\,,\label{eq:betalargeN}
\end{align}
which admit exact solutions as
\begin{align}
    \lambda(\mu)&=\left(\frac{1}{\lambda_0}+\frac{\mathcal{N}}{(4\pi)^2}\ln\left[\frac{\mu}{\mu_0}\right]\right)^{-1}\,,\nonumber\\
    \xi(\mu)&=\frac{420\left(\left[\lambda_0/\lambda(\mu)\right]^{\tilde\Delta/(3\mathcal{N})}-1\right)\lambda(\mu)}{2\tilde\Delta+\left[\tilde\Delta-3(\mathcal{N}+1)\right]\left(\left[\lambda_0/\lambda(\mu)\right]^{\tilde\Delta/(3\mathcal{N})}-1\right)}\,,
\end{align}
where $\tilde\Delta\equiv\sqrt{9\mathcal{N}(\mathcal{N}+2)+79}$. Still in the large-$\mathcal{N}$ limit [meaning as before $\mathcal{N}\gg\mathcal{O}(10)$] and assuming $\mathcal{N}\gg\sqrt{\lambda_0/\lambda}$, these solutions can be simplified to obtain
\begin{equation}
    \xi(\mu)\simeq\frac{70\left[\lambda_0-\lambda(\mu)\right]}{\mathcal{N}}=\frac{70\lambda_0^2\ln(\mu/\mu_0)}{16\pi^2+\lambda_0\mathcal{N}\ln(\mu/\mu_0)}\,.\label{eq:largeNApproxSolxiRun}
\end{equation}
Considering $\lambda_0>0$ from here, we define a 't\,Hooft-like coupling constant,
\begin{equation}
    \lth \equiv\frac{\lambda_0\mathcal{N}}{(4\pi)^2}\,,
\end{equation}
so the solutions to \eqref{eq:betalargeN} read
\begin{equation}
    \lambda(\mu)\simeq\frac{\lambda_0}{1+\lth \ln(\mu/\mu_0)}\,,\qquad\xi(\mu)\simeq\frac{35\lambda_0^2\ln(\mu/\mu_0)}{8\pi^2\left[1+\lth \ln(\mu/\mu_0)\right]}\,.
\end{equation}
From this, we determine that we remain in the regime of validity of the one-loop approximation when $\mu>\mu_0$ so long as $\lth \lesssim 1$, which is to say that $\mathcal{N}\gg 1$ but $\lambda_0\ll 1$ such that their product $\lambda_0\mathcal{N}$ is not larger than $(4\pi)^2$, since then
\begin{equation}
    \left|\frac{\dd\ln\lambda}{\dd\ln\mu}\right|\lesssim\lth \lesssim 1\,.
\end{equation}
Note that if one is in a situation where $\lth \ln(\mu/\mu_0)\ll 1$ (which is to say that $\lambda\approx\lambda_0$), then \eqref{eq:largeNApproxSolxiRun} yields the same leading-order dependence as in \eqref{eq:approxSolBetaVac}. This would not be true, though, when $\lth \ln(\mu/\mu_0)\gtrsim 1$. For our slow-roll inflation, $\lth \ln(\mu/\mu_0)\gtrsim 1$, and thus the approximation in Eq. \eqref{eq:approxSolBetaVac} is not valid.

\subsection{Cosmology}

\subsubsection{Jordan frame}

The classical action of pure quadratic gravity,
\begin{equation}
    S=-\int\dd^4x\,\sqrt{-g}\left(\frac{R^2}{\xi}+\frac{C^2}{2\lambda}\right)\,,
\end{equation}
is scale invariant, but this is expected to be broken at the quantum level. Indeed, the RG flow introduces a dimensionful parameter, $\mu_0$, though it is completely arbitrary at this point. There is also a certain ambiguity in choosing the most appropriate \emph{physical} running scale $\mu$ as a covariant scalar. At the level of flat, isotropic and homogeneous cosmology with only small perturbations on top, one may expect the Hubble scale or derivatives thereof to be reasonably define a physical running scale. Thus, one could choose any scalar invariant of the Riemann tensor, appropriately rescaled to the correct dimension, the simplest of which simply being the Ricci scalar. Hence, we explore the possibility that $\mu=|R|^{1/2}$. Further ignoring the contribution from the Weyl tensor squared for now, this means our (covariant) effective cosmological (quantum) action is
\begin{equation}
    S\simeq -\int\dd^4x\,\sqrt{-g}\,\frac{R^2}{\xi(R)}\simeq -\frac{8\pi^2}{35\lambda_0^2}\int\dd^4x\,\sqrt{-g}\left(\lth +\frac{4}{\ln(R^2/R_0^2)}\right)R^2\,.\label{eq:SofR}
\end{equation}
The above used the approximate solution \eqref{eq:largeNApproxSolxiRun} for the quantum running of $\xi$, and in doing so, we identified the dimensionful running scale $\mu_0$ with $\sqrt{R_0}$ (assuming $R_0>0$), though we use both interchangeably.

\subsubsection{Evolution}

The action \eqref{eq:SofR} corresponds to an $f(R)$ theory of modified gravity with $f(R)=8\pi^2R^2[\lth+4/\ln(R^2/R_0^2)]/(35\lambda_0^2)$. The equation of motion of such a theory follows as
\begin{equation}
    f'(R)R_{\mu\nu}-\frac{1}{2}f(R)g_{\mu\nu}+f''(R)\left(g_{\mu\nu}\Box-\nabla_\mu\nabla_\nu\right)R+f'''(R)\left(g_{\mu\nu}\nabla^\alpha R\nabla_\alpha R-\nabla_\mu R\nabla_\nu R\right)=0\,,
\end{equation}
where $\Box\equiv\nabla^\alpha\nabla_\alpha$ denotes the D'Alembertian and where a prime denotes a derivative with respect to the argument of the function. In a flat Friedmann-Lemaître-Robertson-Walker (FLRW) background with metric $g_{\mu\nu}\dd x^\mu\dd x^\nu=-\dd t^2+a(t)^2\delta_{ij}\dd x^i\dd x^j$, the Ricci scalar is given by $R=6(2H^2+\dot H)$, where $H\equiv\dot a/a$ is the Hubble parameter and where an overdot indicates a derivative with respect to the time $t$. Thus, for most $f(R)$ theories of gravity with $f''(R)\not\equiv 0$, the equation of motion contains two derivatives of $R$, hence up to third-order time derivatives of $H$. Since all nongravitational fields are in their vacuum, the evolution is fully prescribed by the constraint equation, which has one fewer time derivative than the evolution equation, hence it contains up to $\ddot H$. (A time derivative of the constraint equation directly yields the equivalent evolution equation, again because the continuity equation is trivially solved when there is no energy-momentum source.) Thus in FLRW, the dynamical system can be expressed as $\partial_t\vec{P}=\vec{F}(\vec{P})$ in terms of the phase-space vector $\vec{P}=\begin{pmatrix}H\\\dot H\end{pmatrix}$. For a general pure $f(R)$ gravity theory, one has $\vec{F}=\begin{pmatrix}\dot H\\\ddot H(H,\dot H)\end{pmatrix}$, with
\begin{equation}
    \ddot H(H,\dot H)=-\frac{f-6(H^2+\dot H)f'+144H^2\dot Hf''}{36Hf''}\,.
\end{equation}
For our $f(R)$ of interest [recall the action \eqref{eq:SofR}], the phase space trajectories are fully set in terms of the coupling parameter $\lth$ and their scale is set by $R_0=\mu_0^2$.

\begin{figure}[ht]
    \centering
    \includegraphics[scale=0.7]{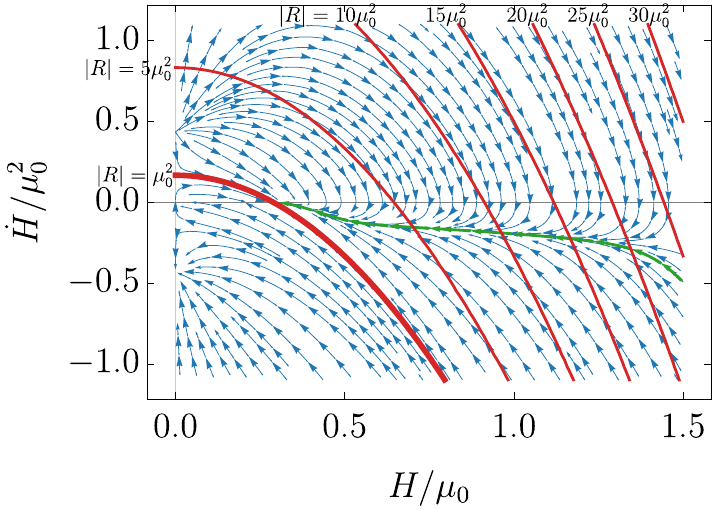}
    \caption{Phase space trajectories corresponding to the FLRW background evolution of the theory \eqref{eq:SofR}. The 't~Hooft-like coupling is set to $\lth=1/10$, and curves of constant $|R|$ are shown in red. The inflationary `attractor' solution is highlighted in green.}
    \label{fig:background}
\end{figure}

In Fig.~\ref{fig:background}, we show the phase space trajectories when setting $\lth=1/10$. We wish to look at trajectories that start in the far UV (where $|R|\gg\mu_0^2$), e.g., at large $H$ and large $-\dot H$ for an expanding cosmology satisfying the null energy condition. (Curves of constant $|R|$ are highlighted in red.) In all cases at large $|R|$, we see that trajectories are attracted toward the green trajectory, which corresponds to `slow-roll' inflation. Indeed, one can compute the slow-roll parameter $\epsilon\equiv-\dot H/H^2$ and find it to be $\ll 1$ along the green trajectory. This confirms that QQG leads to slow-roll inflation under RG flow toward the IR.

We note that in the present case, the thick red line that corresponds to the tachyon divide $|R|=\mu_0^2$ cannot be dynamically crossed. In fact, all trajectories that start from the UV appear to reach a future de Sitter attractor fixed point. It would thus appear that the theory cannot be linked to the IR where GR must emerge. However, if $\lth\equiv\lambda_0\mathcal{N}/(4\pi)^2$ is close to unity, then as $\lambda$ approaches $\lambda_0$ at the point where $\mu=\mu_0$, we naturally enter into the strong coupling regime of the theory. Accordingly, higher loop effects should start coming into play and change the late-time, IR picture. While the big picture is expected to be the same (evolution toward the tachyon-free IR region of quadratic gravity, emergence of GR, and reheating into a hot radiation-filled universe), the actual dynamics for how this is achieved deserves to be further studied. Another thing to keep in mind is that Fig.~\ref{fig:background} changes if another choice is made for the RG scale $\mu$. We explored, for instance, the possibility that $\mu$ is given by the fourth root of the absolute value of the Kretschmann scalar invariant. In such a situation, we found that the slow-roll attractor trajectory in the UV remains (since both the Kretschmann scalar and the Ricci scalar are mainly controlled by $H$ when $-\dot H\ll H^2$). However, the late-time behavior of the inflationary trajectories were different, as they could cross the tachyon divide and reach an IR fixed point. How different choices of scalar invariants for $\mu$ affect perturbations would also have to be explored in future work.

\subsubsection{Einstein frame}

Upon introducing the dimensionless auxiliary variable $\psi=R/R_0$, the action expressed as $S=-\int\dd^4x\,\sqrt{-g}\,f(R)$ with $f(R)=R^2/\xi(R)$ becomes one with $f(\psi)=8\pi^2\mu_0^4\psi^2(\lth +4/\ln[\psi^2])/(35\lambda_0^2)$. Defining $\Omega^2(\psi)\equiv 2f'(\psi)/\mu_0^4$ and $U(\psi)\equiv\psi f'(\psi)-f(\psi)$, the action can be recast as
\begin{equation}
    S=-\int\dd^4x\,\sqrt{-g}\left(\frac{\mu_0^2}{2}\Omega^2(\psi)R-U(\psi)\right)\,,\label{eq:actionpsi}
\end{equation}
where the expressions are found to be
\begin{align}
    \Omega^2(\psi)&=\frac{32\pi^2\psi}{35\lambda_0^2}\left[\lth +4\left(\frac{1}{\ln[\psi^2]}-\frac{1}{\ln^2[\psi^2]}\right)\right]\,,\nonumber\\
    U(\psi)&=\frac{8\pi^2\mu_0^4\psi^2}{35\lambda_0^2}\left[\lth +4\left(\frac{1}{\ln[\psi^2]}-\frac{2}{\ln^2[\psi^2]}\right)\right]\,.
\end{align}
With those, one can verify that varying \eqref{eq:actionpsi} with respect to $\psi$ indeed yields $\psi=R/\mu_0^2$, hence $(\mu_0^2/2)\Omega^2(\psi)R-U(\psi)=f(R)$.

\begin{figure}[ht]
    \centering
    \includegraphics[width=0.47\linewidth]{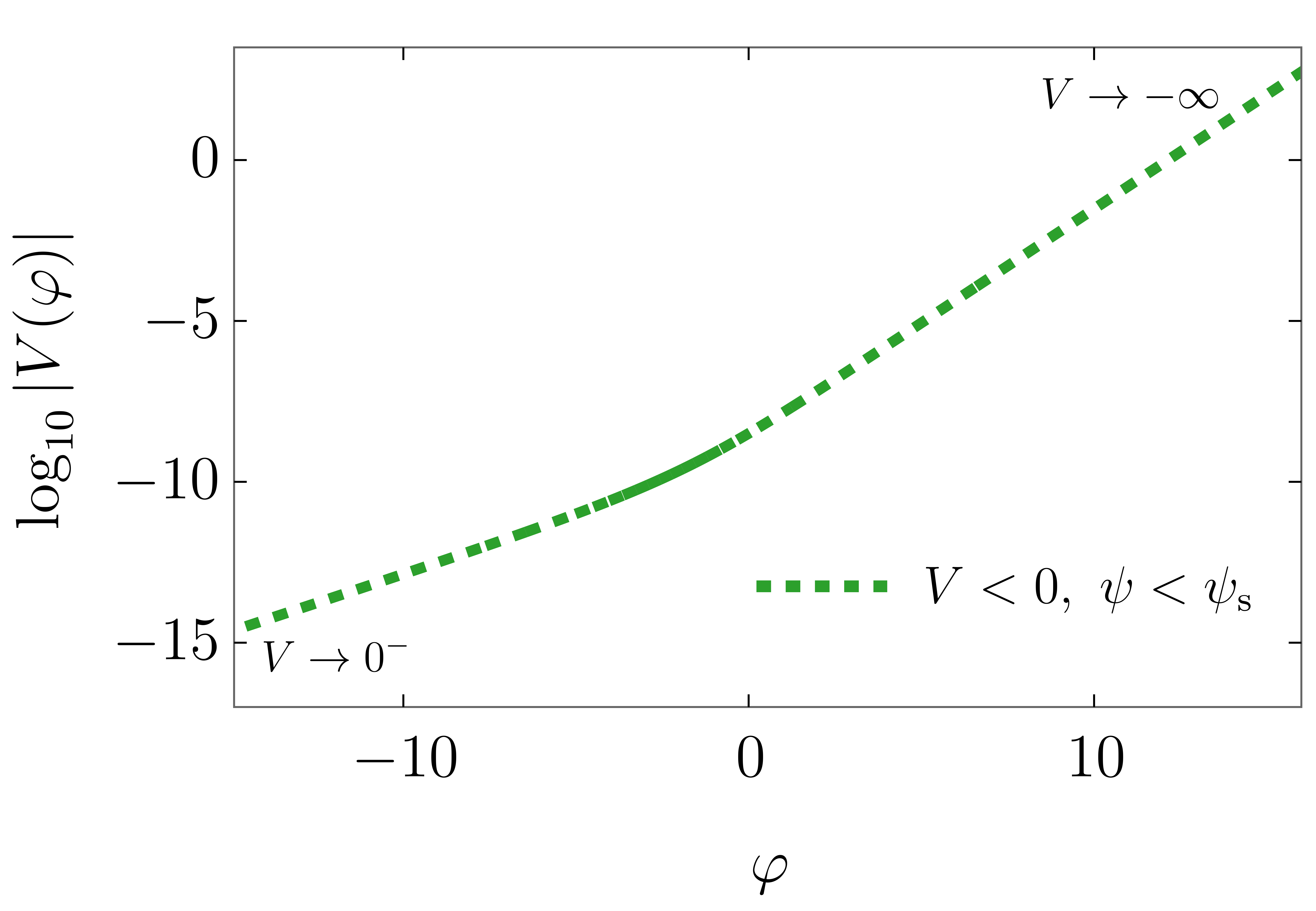}\hspace*{0.05\linewidth}\includegraphics[width=0.47\linewidth]{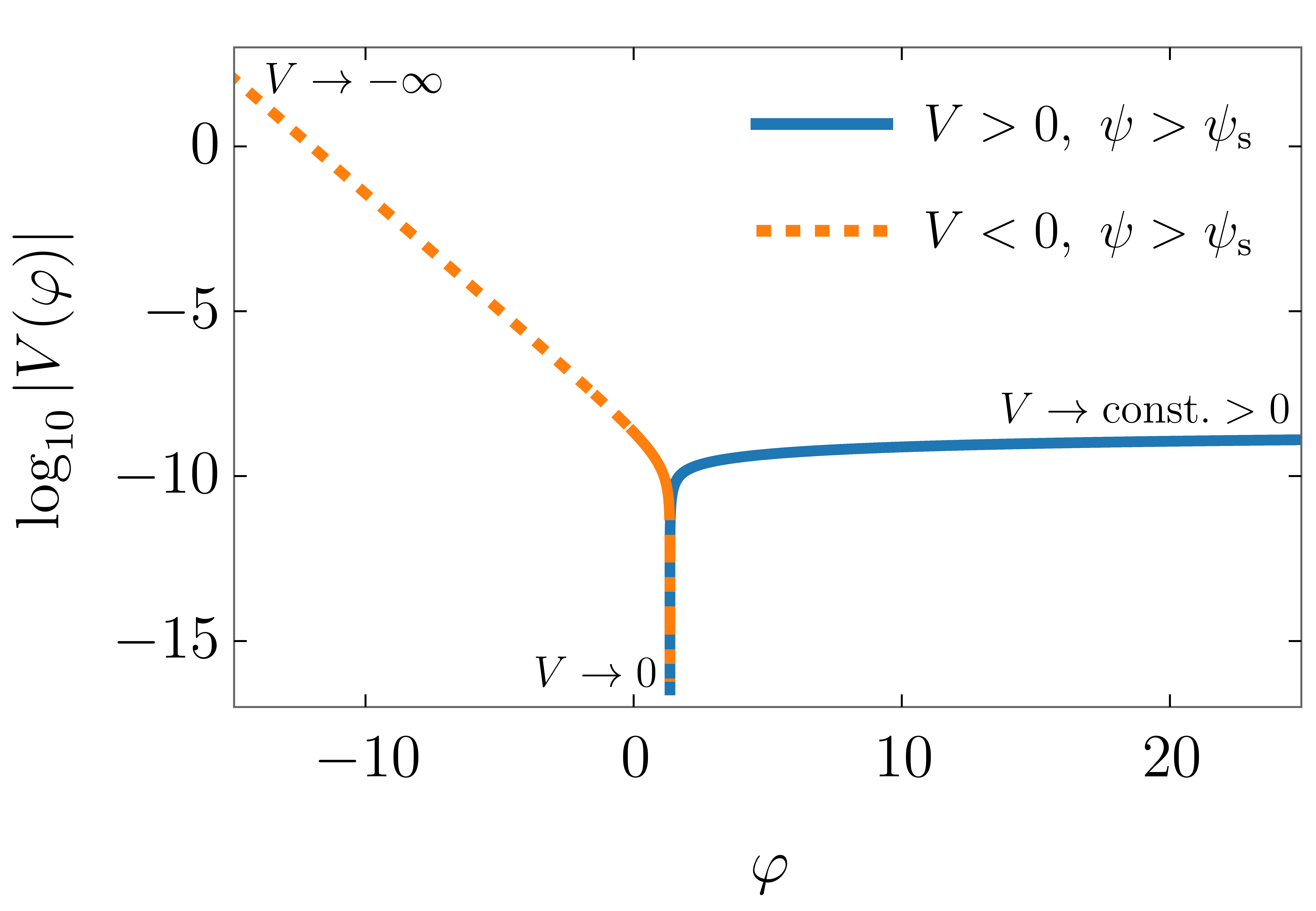}
    \caption{Full Einstein-frame potential, parametrically given by \eqref{eq:varphisol} and \eqref{eq:Vtilde} as a function of the dimensionless RG scale $\psi$. The absolute value of the potential is plotted on a logarithmic scale, so we distinguish between positive and negative branches of the potential by using solid and dashed lines, respectively. The right panel corresponds to the branch of the potential when $\psi>\psi_\mathrm{s}$, and vice versa, the the left panel shows the potential when $\psi<\psi_\mathrm{s}$.}
    \label{fig:potential}
\end{figure}

We then perform a change of variable for the metric,
\begin{equation}
    \tilde g_{\mu\nu}=\Omega^2(\psi)g_{\mu\nu}\,,\label{eq:conftransf}
\end{equation}
provided $\Omega^2(\psi)>0$, which is to say that $\ln[\psi^2]>\ln[\psi_\mathrm{s}^2]\equiv2/\left(1+\sqrt{1+\lth }\right)$. Here, $\psi_\mathrm{s}$ determines the value at which the conformal transformation \eqref{eq:conftransf} is singular. The action then transforms, up to boundary terms, to
\begin{equation}
    S=-\int\dd^4x\,\sqrt{-\tilde g}\left(\frac{\mu_0^2}{2}\tilde R-\mathcal{\tilde K}(\psi)\tilde g^{\mu\nu}\nabla_\mu\psi\nabla_\nu\psi-\tilde V(\psi)\right)\,,\label{eq:EFa1}
\end{equation}
provided
\begin{align}
    \mathcal{\tilde K}(\psi)&=\frac{3\mu_0^2}{4}\left(\frac{\dd\ln[\Omega^{2}(\psi)]}{\dd\psi}\right)^2=\frac{3\mu_0^2}{4\psi^2\ln^2[\psi^2]}\left(\frac{\lth  \ln^3[\psi^2]+4\big(\ln^2[\psi^2]-3\ln[\psi^2]+4\big)}{\lth  \ln^2[\psi^2]+4\big(\ln[\psi^2]-1\big)}\right)^2\,,\nonumber\\
    \tilde V(\psi)&=\frac{U(\psi)}{\Omega^4(\psi)}=\frac{35\mu_0^4\lambda_0^2\ln^2[\psi^2]\Big(\lth  \ln^2[\psi^2]+4\big(\ln[\psi^2]-2\big)\Big)}{128\pi^2\Big(\lth  \ln^2[\psi^2]+4\big(\ln[\psi^2]-1\big)\Big)^2}\,.\label{eq:Vtilde}
\end{align}
We can then write the resulting `Einstein-frame action' with a canonically normalized scalar field $\varphi$ and potential $V(\varphi)$ --- we drop tildes when it is clear that we are in the Einstein frame --- where $\dd\varphi/\dd\psi=\pm\sqrt{2\mathcal{\tilde K}(\psi)}$, hence
\begin{equation}
    \varphi(\psi)=\pm\mu_0\sqrtb{\frac{3}{2}}\ln\left[\psi\left|\lth +4\left(\frac{1}{\ln[\psi^2]}-\frac{1}{\ln^2[\psi^2]}\right)\right|\right]\,.\label{eq:varphisol}
\end{equation}
This cannot be analytically inverted, but in the limit where $\ln[\psi^2]\gg 2\left(1+\sqrt{1-\lth }\right)/\lth $, we get $\psi\simeq\lth ^{-1}\exp(\sqrtb{2/3}\,\varphi/\mu_0)$, and thus,
\begin{equation}
    V(\varphi)\simeq\frac{35\lambda_0^2\mu_0^4}{128\pi^2\lth }\left(1-\frac{\sqrtb{6}\mu_0}{\lth \varphi}\right)\,.\label{eq:Vapprox}
\end{equation}
This potential captures well the early time, slow-roll dynamics. The full potential is shown in Fig.~\ref{fig:potential}, and \eqref{eq:Vapprox} amounts to the blue part of the potential that is on the far right of the right panel.

While slow-roll inflation ends as the blue curve steepens and as the potential approaches $0$, we can see that the full potential has interesting additional features: first, the potential crosses $0$ when $\lth\ln^2[\psi^2]+4(\ln[\psi^2]-2)=0$, after which it becomes negative (dashed curves in Fig.~\ref{fig:potential}). Then, as $\psi$ approaches $\psi_\mathrm{s}$ from above, the potential becomes infinitely negative at the same time as $\varphi\to-\infty$. Using \eqref{eq:Vtilde} and \eqref{eq:varphisol}, we can also plot the other branch of the potential, `past the singularity' in the conformal transformation \eqref{eq:conftransf}. There (left panel in Fig.~\ref{fig:potential}), the potential `starts' from $-\infty$ when $\varphi\to\infty$ and approaches $0$ from below as $\varphi\to-\infty$. In all cases, $\varphi$ undergoes kination after inflation. This might appear rather different than the Jordan frame picture shown earlier (where the late-time behavior was de Sitter). However, we recall that the background scale factors in both frames are nonequivalent, though they are related through the conformal relation $\tilde a^2(\tilde t)=\Omega^2(\psi)a^2(t)$ following \eqref{eq:conftransf}. Still, today's observables should be the same when properly calculated (e.g., \cite{Anselmi:2020lpp}).

At this point, let us remark that our Einstein-frame action written as \eqref{eq:EFa1} with \eqref{eq:Vtilde} has $\mu_0$ as its only dimensionful parameter since this is the only scale that appears once quantum running breaks scale invariance. This scale is arbitrary, though, and to properly `match' \eqref{eq:EFa1} with GR, a natural choice is to take $\mu_0=\mpl$. Let us stress, however, that the Planck mass and GR are not yet present in QQG in the UV --- they need to emerge in the IR in QQG's strongly coupled regime. Still, without a proper understanding of GR's emergence, we can take $\mu_0=\mpl$, and we can think of QQG and GR as living on two sides of a UV/IR matching surface. Putting this consideration aside, in what follows we will see that the choice of $\mu_0$ is irrelevant in the calculation of observable quantities.

An advantage of the Einstein frame is the applicability of its slow-roll parameters to compute cosmological perturbations and their corresponding observables, such as the scalar spectral index $n_\mathrm{s}$, the scalar amplitude $A_\mathrm{s}$, and the tensor-to-scalar ratio $r$. The first slow-roll parameter of the potential is found to be
\begin{equation}
    \epsilon_V=\frac{\mu_0^2}{2}\left(\frac{V'(\varphi)}{V(\varphi)}\right)^2=\frac{\mu_0^2}{4\mathcal{\tilde K}(\psi)}\left(\frac{\tilde V'(\psi)}{\tilde V(\psi)}\right)^2=\frac{1}{3}\left(\frac{8}{\lth  \ln^2[\psi^2]+4\big(\ln[\psi^2]-2\big)}\right)^2\,,
\end{equation}
and already, this tells us that inflation approximately ends at some value of $\psi$ that we call $\psi_\mathrm{e}$ when
\begin{equation}
    \epsilon_V=1\qquad\implies\qquad\ln[\psi_\mathrm{e}^2]=\frac{2}{\lth }\left(-1+\sqrt{1+2\left(1+\frac{1}{\sqrt{3}}\right)\lth }\right)\,.\label{eq:endInf}
\end{equation}
The second slow-roll parameter of the potential is found to be
\begin{align}
    \eta_V=\mu_0^2\frac{V''(\varphi)}{V(\varphi)}&=\frac{\mu_0^2}{\tilde V(\psi)\sqrt{2\mathcal{\tilde K}(\psi)}}\frac{\dd}{\dd\psi}\left(\frac{\tilde V'(\psi)}{\sqrt{2\mathcal{\tilde K}(\psi)}}\right)\nonumber\\
    &=-\frac{64\left(\lth  \ln^2[\psi^2]+4\right)}{3\Big(\lth  \ln^2[\psi^2]+4\big(\ln[\psi^2]-2\big)\Big)\Big(\lth  \ln^3[\psi^2]+4\big(\ln^2[\psi^2]-3\ln[\psi^2]+4\big)\Big)}\,,
\end{align}
and thus, we find the spectral index
\begin{align}
    n_\mathrm{s}=1+2\eta_V-6\epsilon_V=&~1-\frac{128}{\Big(\lth  \ln^2[\psi^2]+4\big(\ln[\psi^2]-2\big)\Big)^2}\nonumber\\
    &-\frac{128\left(\lth  \ln^2[\psi^2]+4\right)}{3\Big(\lth  \ln^2[\psi^2]+4\big(\ln[\psi^2]-2\big)\Big)\Big(\lth  \ln^3[\psi^2]+4\big(\ln^2[\psi^2]-3\ln[\psi^2]+4\big)\Big)}\,,
\end{align}
the tensor-to-scalar ratio
\begin{equation}
    r=16\epsilon_V=\frac{1}{3}\left(\frac{32}{\lth  \ln^2[\psi^2]+4\big(\ln[\psi^2]-2\big)}\right)^2\,,
\end{equation}
and the scalar amplitude
\begin{equation}
    A_\mathrm{s}=\frac{V(\varphi)^3}{12\pi^2\mu_0^6V'(\varphi)^2}=\frac{\mathcal{\tilde K}(\psi)\tilde V(\psi)^3}{6\pi^2\mu_0^6\tilde V'(\psi)^2}=\frac{35\lambda_0^2\ln^2[\psi^2]\Big(\lth  \ln^2[\psi^2]+4\big(\ln[\psi^2]-2\big)\Big)^3}{\left(256\pi^2\Big[\lth  \ln^2[\psi^2]+4\big(\ln[\psi^2]-1\big)\Big]\right)^2}\,.
\end{equation}
While all of these are functions of the dimensionless RG scale $\psi$, it is more useful to reexpress them in terms of the inflationary $e$-folding number $N$, which we define to be the number of $e$-folds before the end of inflation. Using \eqref{eq:endInf}, we find
\begin{align}
    N(\psi)=&~\frac{1}{\mu_0^2}\int_{\varphi_\mathrm{e}}^{\varphi}\dd\tilde\varphi\,\frac{U(\tilde\varphi)}{U'(\tilde\varphi)}=\frac{2}{\mu_0^2}\int_{\psi_\mathrm{e}}^{\psi}\dd\tilde\psi\,\frac{\mathcal{\tilde K}(\tilde\psi)\tilde V(\tilde\psi)}{\tilde V'(\tilde\psi)}\nonumber\\
    =&~\frac{\lth \ln^3[\psi^2]}{32}+\frac{3\ln^2[\psi^2]}{16}-\frac{3\ln[\psi^2]}{2}-\frac{3}{4}\ln\left(\frac{\lth }{\ln^2[\psi^2]}+\frac{4}{\ln^3[\psi^2]}-\frac{4}{\ln^4[\psi^2]}\right)\nonumber\\
    &-\frac{3+(15-\sqrt{3})\lth }{6\lth ^2}\sqrtb{1+2\left(1+\frac{1}{\sqrt{3}}\right)\lth }+\frac{1}{2\lth ^2}+\frac{3}{\lth }-\frac{3}{4}\ln\left(\frac{(3+2\sqrt{3})\lth ^4}{12\left(\sqrtb{1+2\left(1+1/\sqrtb{3}\right)\lth }-1\right)^4}\right)\label{eq:Nofpsifull}\\
    \approx&~\frac{\lth \ln^3[\psi^2]}{32}+\frac{3\ln^2[\psi^2]}{16}-\frac{3\ln[\psi^2]}{2}\,,\nonumber
\end{align}
where the last approximation assumes $\psi$ is large.
In fact, sufficiently early during inflation (i.e., when $N$ is sufficiently large), one can estimate $N\sim(\lth /32)\ln^3[\psi^2]\implies\ln[\psi^2]\sim 2(4N/\lth )^{1/3}$. From this, we can find rough estimates of the observables as
\begin{equation}
    n_\mathrm{s}\sim 1-\frac{4}{3N}\,,\qquad r\sim\frac{8}{3}\left(\frac{2}{\lth ^2N^4}\right)^{1/3}\,,\qquad A_\mathrm{s}\sim\frac{35\lambda_0^2}{512\pi^4}\left(\frac{N^4}{2\lth }\right)^{1/3}\,.
\end{equation}

\subsubsection{Reheating}

In the main text, we assumed the number of $e$-folds $N$ between a mode's horizon exit and the end of inflation to be in the standard $50$ to $60$ range. In general, this is determined by \cite{Liddle:2003as,Martin:2010kz,Planck:2018jri}
\begin{equation}
    N\approx 67-\ln\left(\frac{k_\star}{a_0H_0}\right)+\frac{1}{4}\ln\left(\frac{V_\star^{1/2}}{\mpl^2\rho_\mathrm{end}}\right)+\frac{1-3w}{12(1+w)}\ln\left(\frac{\rho_\mathrm{reh}}{\rho_\mathrm{end}}\right)-\frac{1}{12}\ln g_\mathrm{reh}\,,\label{eq:Nreh}
\end{equation}
where $k_\star$ is the mode's comoving wavenumber, $a_0H_0$ is the conformal Hubble scale today, $V_\star$ is the value of the inflationary potential when the mode $k_\star$ exits the horizon, $\rho_\mathrm{end}$ is the energy density at the end of inflation, $w$ is the post-inflationary equation-of-state (EoS) parameter until the end of reheating, $\rho_\mathrm{reh}$ is the energy density once the universe has thermalized (so at the end of reheating), and $g_\mathrm{reh}$ is the number of effective bosonic degrees of freedom at the scale $\rho_\mathrm{reh}$.

\begin{figure}[ht]
    \centering
    \includegraphics[width=0.5\linewidth]{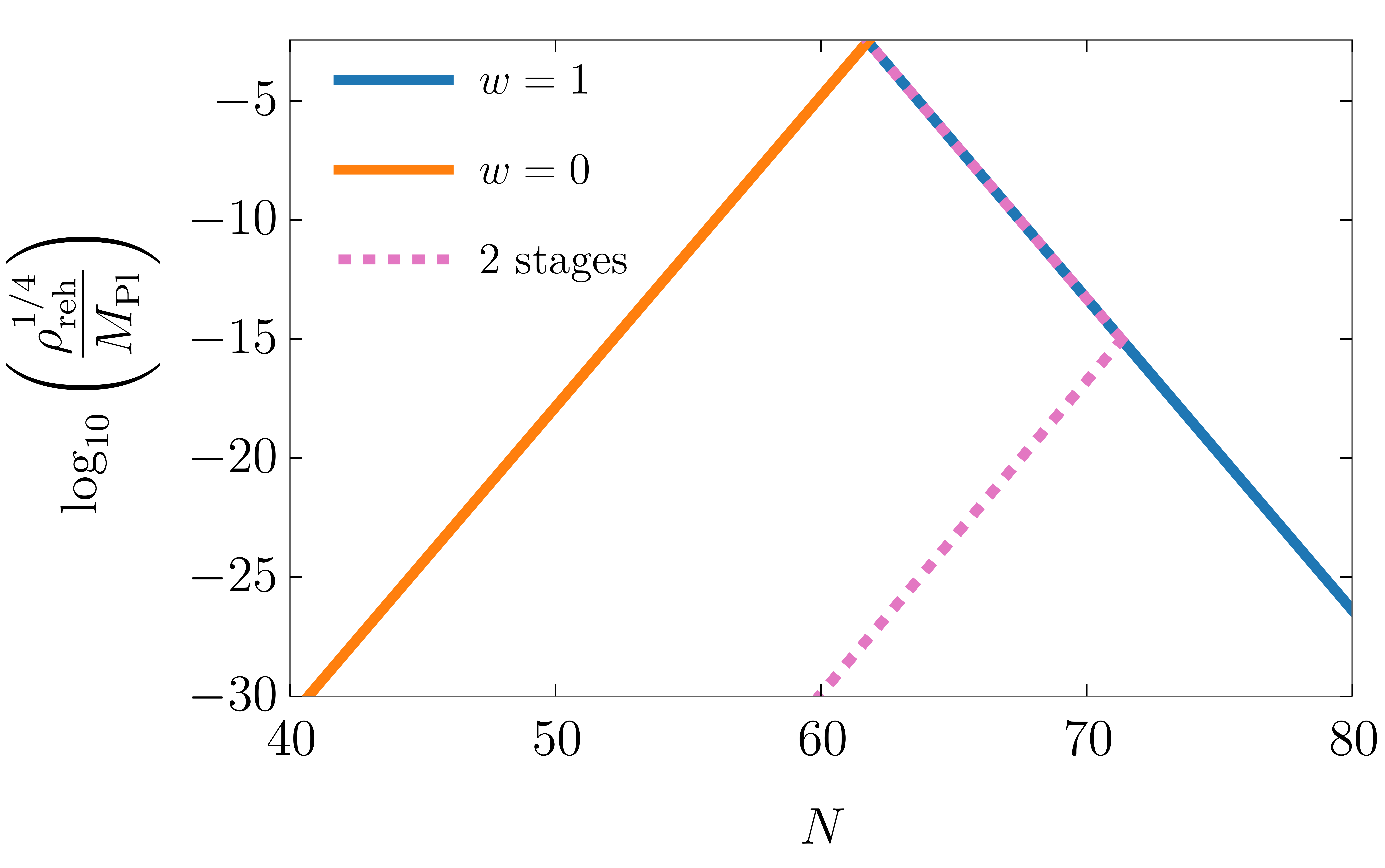}
    \caption{Plot showing the reheating energy scale $\rho_\mathrm{reh}^{1/4}$ in Planck units versus the number of inflationary $e$-folds $N$ needed between horizon exit and the end of inflation. The different lines present different post-inflationary histories: kination until reheating occurs instantaneously (the EoS parameter is $w=1$ throughout) in blue; no kination but slow reheating with EoS parameter $w=0$ in orange; and a 2-stage history in dashed pink with kination ($w=1$) for a while and then reheating (with $w=0$) after that.}
    \label{fig:reheating}
\end{figure}

Setting standard values for $k_\star$, $a_0H_0$, and $g_\mathrm{reh}$, we can use Eqs.~\eqref{eq:Nofpsifull}, \eqref{eq:Vtilde}, and \eqref{eq:endInf} to compute $N$ in terms of $V_\star$ and $\rho_\mathrm{end}\approx V(\psi_\mathrm{e})$ for our model. Then, using Eq.~\eqref{eq:Nreh}, we can plot the reheating energy scale $\rho_\mathrm{reh}^{1/4}$ as a function of $N$ for different values of $w$; see Fig.~\ref{fig:reheating}. From the plot, we note that if there is no kination phase after inflation and if reheating is instantaneous, then $N\sim 62$ and the thermalization scale is about the same scale as the inflationary one near the Grand Unified Theory scale. More realistically, our model indicates that there will be some kination phase after inflation, though it is hard to compute how long it would last without a proper reheating model (which would fall in the strongly coupled regime of the theory). Still, we see from the blue line of Fig.~\ref{fig:reheating} that the longer the kination phase ($w=1$), the smaller the reheating scale, and the larger $N$ would need to be if reheating is to be instantaneous. This would push $n_\mathrm{s}$ to larger values and put the model in tension with data. If reheating is not instantaneous, but if instead it lasts some time with EoS $w=0$ for instance (see orange line of Fig.~\ref{fig:reheating}), then this would require a smaller value of $N$. A more realistic post-inflationary history could be a 2-stage process with kination first and then reheating afterwards; the pink dashed curves emulates such a scenario. Assuming $N\approx 60$ appears to be reasonable in that sense, though again, we must stress that a proper assessment of the duration and EoS of kination and reheating is currently out of reach and would require a better understanding of the phase transition from QQG to GR, which occurs in the strongly coupled regime of QQG. This is the reason behind simply assuming $N\sim 50-60$ at this point, in line with the common practice in inflationary model building.

\end{document}